\documentclass[aps,showpacs,showkeys,twocolumn,preprintnumbers,amsmath,amssymb]{revtex4}

\UseRawInputEncoding
\usepackage{hyperref}
\hypersetup{colorlinks=true,
            linkcolor=magenta,
            filecolor=black,
            urlcolor=magenta,
            citecolor=magenta
           }

\usepackage{dcolumn}

\usepackage{float}
\usepackage{graphicx}

\usepackage{epstopdf}
\usepackage{graphicx}
\usepackage{epstopdf}
\usepackage{color}
\usepackage{amsmath,amssymb,amsfonts}

\makeatletter

\newcommand{\Rmnum}[1]{\expandafter\@slowromancap\romannumeral #1@}
\makeatother

\begin{document}
\baselineskip=0.5 cm

\title[Article Title]{Polarization Signatures of Sgr A* with Tilted Hybrid Magnetic Fields}

\author{Yushu Feng, Hao Yin, Songbai Chen\footnote{Corresponding author: csb3752@hunnu.edu.cn}, Jiliang Jing\footnote{jljing@hunnu.edu.cn}}

\affiliation{Department of Physics, Institute of Interdisciplinary Studies, Hunan Research Center of the Basic Discipline for Quantum Effects and Quantum Technologies, Key Laboratory of Low Dimensional Quantum Structures and Quantum Control of Ministry of Education, Synergetic Innovation Center for Quantum Effects and Applications, Hunan Normal University,  Changsha, Hunan 410081, People's Republic of China.}
%%%%%%%%%%%%%%%%%%%%%%%%%%%%%%%%%%%%%%%%%%%%%%%%%%%%%%%%%%%%%%%%%%%%%%%%%%%%
\begin{abstract}
\baselineskip=0.4 cm
The magnetic field geometry in the near-horizon region is very important in shaping polarimetric signatures of black hole images. With a hybrid magnetic field consisting of an axisymmetric split-monopole field and a tilted uniform external field, we  investigate  the effects of external magnetic field inclination on the polarimetric signatures of Sgr A* by employing a stationary semi-analytical radiatively inefficient accretion flow (RIAF) model and general relativistic polarized radiative transfer simulations.
Our results show that the external magnetic field tilt significantly modifies the electric vector position angle (EVPA) pattern and brightness distribution. The induced non-axisymmetry from the tilted magnetic filed component renders polarization morphology dependent on the observer's azimuthal angle, which is a novel behavior absent in purely axisymmetric magnetic field configurations. In combination with Event Horizon Telescope (EHT) data, we analyse the possibility of such hybrid magnetic fields around Sgr A*. 
\end{abstract}

%%%%%%%%%%%%%%%%%%%%%%%%%%%%%%%%%%%%%%%%%%%%%%%%%%%%%%%%%%%%%%%%%%%%%%%%%%%%
\pacs{04.70.-s, 98.62.Mw, 97.60.Lf}
\keywords{Tilted Hybrid Magnetic Fields; Polarization Signatures;  radiatively inefficient accretion flow;}
\maketitle
%%%%%%%%%%%%%%%%%%%%%%%%%%%%%%%%%%%%%%%%%%%%%%%%%%%%%%%%%%%%%%%%%%%%%%%%%%%%
\section{Introduction}\label{sec:intro}
Recently, the Event Horizon Telescope (EHT) has sequentially released horizon-scale polarimetric images of M87* and Sgr A*, which provides a new avenue for exploring the matter distribution,  electromagnetic emission and associated dynamics in the near-horizon regions \cite{1,2,3,37,38,39,5,6,7,8,42,47,48,49,50,51,52,53,4}. Magnetic fields affect the intrinsic synchrotron emission and radiative transfer in black hole accretion flows, and are a key ingredient in the formation of black hole images.
In images of Sgr A*, the rapid polarimetric variability and prominent local asymmetries within its large-scale spiral polarization structure indicate an extremely short dynamical timescale in the vicinity of the black hole together with complicated Faraday rotation \cite{4,9,10,11}. This also implies that the magnetic field configuration around the black hole is highly complex.

The influence of magnetic fields on black hole images has been extensively investigated \cite{10,11,12,13,14,15,16,17,18,19,20,21,22,46,51,53}. 
By confronting general relativistic magnetohydrodynamic (GRMHD) simulations \cite{36} with observational constraints, the EHT Collaboration indicates that the polarization signatures of both M87* and Sgr A* strongly favor a magnetically arrested disk (MAD) accretion state \cite{10,11,12,13,14}. Moreover, using semi-analytical radiatively inefficient accretion flow (RIAF) model, the effects of magnetic field geometries on linear-polarized \cite{15,16,17,18,19} and circular-polarized images of black holes \cite{20} have been investigated. These investigations indicate that the magnetic field configuration around black holes plays an important role in determining polarimetric images. Most existing works assume the magnetic field configuration surrounding the black hole to be symmetric about its spin axis. However, such perfect alignment may not generally hold under astrophysical conditions. Progenitor systems producing stellar-mass black holes, including massive-star core collapse and binary neutron star mergers, typically contain complex, turbulent, differentially rotating magnetized plasmas \cite{Obergaulinger:2008pb,Obergaulinger:2017qno,Kiuchi:2014hja,Kiuchi:2015sga}. In these processes, due to asymmetric accretion, off-axis jet launching, or misalignment between the angular momentum of the collapsing core and the large-scale ambient field, the magnetic field inherited by the black hole may become tilted relative to its spin axis \cite{Proga:2003ap,Liska:2017alm}. Moreover, inclined magnetic fields are also anticipated for isolated black holes accreting material from the interstellar medium \cite{El-Badry:2023pah,Barkov:2012sj,Kin:2025axi,Figueiredo:2025xbo}. Nevertheless, the impacts of inclined magnetic fields on black hole images remain an open issue.

Accretion flows constitute the physical light source that generates black hole images in astrophysical environments. Fully describing accretion flow dynamics generally relies on high-precision GRMHD simulations.
Although GRMHD simulations excel at capturing the non-linear dynamic evolution of accretion disks \cite{16,17,18}, the associated high computational cost  poses enormous challenges for systematically exploring the vast parameter space. 
Furthermore, within the ideal GRMHD framework,  both the self-consistent co-evolution of magnetic fields and plasma and the rapid disruption of the initial magnetic topology by magnetic reconnection and turbulence make it exceedingly difficult to independently probe the impact of magnetic structure on observable physical quantities \cite{19}.
For the current EHT targets M87* and Sgr A*, the
surrounding hot plasma is interpreted as belonging to a radiatively inefficient accretion flow (RIAF), a structure typically hot, geometrically thick, and optically thin at the observing frequency of 230 GHz. Therefore, the semi-analytic radiatively inefficient accretion flow (RIAF) model capturing the features of magnetized accretion flows near black holes has been adopted to probe horizon-scale images of various black holes \cite{20,21,22,46,51,53}. 
As a complementary approach, globally stationary semi-analytical RIAF models enable the effective decoupling of the magnetic configuration from non-linear fluid dynamic effects, allowing us to conduct parameterized studies on how diverse magnetic field geometries affect polarimetric signatures. 

Owing to the presence of not only intrinsic magnetic fields produced by strong-gravity accretion flows but also external background fields permeating the black hole from its large-scale surroundings, the interplay and superposition of these internal and external fields severely complicate the near-horizon magnetic structure, fundamentally breaking the global axisymmetry of the system.
Motivated by these considerations, in this work, we will study effects of such inclined magnetic fields on black hole images. We 
first construct a tilted hybrid magnetic field configuration: an axisymmetric split-monopole field serving as the intrinsic magnetic component, linearly superimposed with an external uniform magnetic field with an inclination angle. Then, we  systematically investigate  the effects of external magnetic field inclination,  observer's inclination angle and  azimuthal angle
on the polarimetric signatures of Sgr A*, employing a stationary semi-analytical RIAF model. The remainder of this paper is organized as follows. Sect. II details the RIAF model and its numerical setup, along with the construction of the tilted hybrid magnetic field configuration. Sect. III presents the polarimetric images and the corresponding observables. Sect.IV compares the model predictions with the EHT observational constraints. Finally, Sect.V summarizes the main conclusions.

\section{MODEL AND NUMERICAL SETUP}
\label{sec:cp_mechanism}

In this section, we introduce the semi-analytical RIAF model, the tilted hybrid magnetic field configuration, and the polarimetric observables employed throughout this work.

\subsection{Semi-analytical  RIAF model}

It is well known that the hot plasmas surrounding Sgr A* are considered to be part of a RIAF, we here adopt a semi-analytical optically thin RIAF model \cite{31} around a Schwarzschild black hole to perform general relativistic ray-tracing numerical simulations.  In this semi-analytical RIAF model, the electron number density $n_e$ and temperature $T_e$ are assumed to obey power-law distributions as functions of radius, whose forms can be respectively given by \cite{22,23,24,25,26,45,48,27,52} 
\begin{align}
n_e &= n_{e,0}\left(\frac{r}{r_g}\right)^{-\delta}
\exp\left[-\frac{1}{2}(H\tan\theta)^{-2}\right], \\
T_e &= T_{e,0}\left(\frac{r}{r_g}\right)^{-\gamma},
\end{align}
Here, $r_g = GM_{\rm BH}/c^2$ is the gravitational radius and $H$ parametrizes the geometric thickness of the disk.  $\delta$ and $\gamma$ are the power-law indices for electron density and electron temperature respectively. From the observations of Sgr A* and previous studies of RIAFs \cite{28,29,30}, these two indices can be set to $\delta = 1.1$ and $\gamma = 0.84$, respectively. The electron temperature  and  the electron number density are set to ensure that the total flux density at 230 GHz matches the observed value of $\mathcal{I}_{\rm tot} \approx 2.4\,\mathrm{Jy}$ \cite{40,41}. 

The magnetic field strength is determined by imposing a constant magnetization:
\begin{align}
\sigma = \frac{B^2/4\pi}{m_p c^2 n_e},
\end{align}
where $B$ and $m_p$ denote the magnetic field magnitude and proton mass. We here adopt a low magnetization of $\sigma = 0.01$, corresponding to a weakly magnetized disk in the SANE regime \cite{16,27}. This also ensures that in our simulations the magnetic field strength at the equatorial ring ($r=3r_g$,$\theta=\pi/2$) is approximately $20-30$ G,
which is consistent with  the counterpart of Sgr A* from astronomical observations \cite{10}. 

The four-velocity field of the accretion flow is modeled by interpolating between Keplerian orbital motion and
geodesic free fall \cite{20,25,31}, which introduces $\kappa_{\rm K}$ and $\kappa_{\rm ff}$ as factors to 
control the weights of Keplerian orbital motion and geodesic free fall components, respectively.
\begin{align}
u^r &= u^r_{\rm K} + \kappa_{\rm ff}\left(u^r_{\rm ff} - u^r_{\rm K}\right), \\
\Omega &= \Omega_{\rm K} + \left(1 - \kappa_{\rm K}\right)\left(\Omega_{\rm ff} - \Omega_{\rm K}\right),
\end{align}
where $u^r$ and $\Omega = u^\phi/u^t$ denote the radial four-velocity component and the angular velocity, respectively. We adopt $(\kappa_{\rm ff}, \kappa_{\rm K}) = (0.5, 0.5)$ as our fiducial parameters.
\begin{table}[t]
\centering
\begin{tabular}{ccc}
\hline\hline
\textbf{Parameter} & \textbf{Value} & \textbf{Parameter Description} \\
\hline
$M_{\rm BH}$ & $4.3 \times 10^6 M_\odot$ & Black hole mass \\
$D_s$ & $8.3 \times 10^3\,\mathrm{pc}$ & Distance to the source \\
$\delta$ & $1.1$ & $n_e$ power law index \\
$\gamma$ & $0.84$ & $T_e$ power law index \\
$H$ & $0.3$ & Disk thickness \\
$i$ & $30^\circ$ & Observer inclination angle \\
$\phi$ & $0^\circ$ & Observer azimuthal angle  \\
$\kappa_{\rm K}$ & $0.5$ & Keplerian parameter \\
$\kappa_{\rm ff}$ & $0.5$ & Radial infall parameter \\
$\sigma$ & $0.01$ & Magnetization \\
$n_{e,0}$ & $10^6$--$10^7\,\mathrm{cm}^{-3}$ & \begin{tabular}[c]{@{}c@{}}The normalization of\\electron distribution\end{tabular} \\
$T_{e,0}$ & $5 \times 10^{11}\,\mathrm{K}$ & \begin{tabular}[c]{@{}c@{}}The normalization \\ of temperature\end{tabular}  \\
$\nu_{\rm obs}$ & $230\,\mathrm{GHz}$ & Observing frequency \\
$B_0$ & $1$ & Hybrid field parameter \\
FOV & $200\,\mu\mathrm{as}$ & Field of view \\
$N_X \times N_Y$ & $200 \times 200\,\mathrm{px}$ & Number of pixels \\
\hline\hline
\end{tabular}
\caption{Fiducial Parameters for RIAF}
\label{tab:riaf-parameters}
\end{table}
Other fiducial parameters are summarized in \hyperref[tab:riaf-parameters]{Table~\ref*{tab:riaf-parameters}}.

\subsection{Tilted hybrid magnetic field configuration}
In real astrophysical environments, the magnetic field surrounding the black hole is not necessarily axisymmetric. This is because progenitor systems leading to black hole formation typically involve complex, turbulent, differentially rotating magnetized fluids. Asymmetric accretion, off-axis jet launching, or misalignment between the angular momentum of the collapsing core and the large-scale ambient field also cause the magnetic field inherited by the black hole to be tilted relative to its spin axis. Moreover, isolated black holes accreting from the interstellar medium are also expected to host inclined magnetic fields. In this work, we assume that a supermassive black hole is embedded in the tilted hybrid magnetic field configuration. The tilted component may correspond to large-scale background magnetic fields originating in the interstellar medium, which can be reasonably approximated as uniform over a localized finite spatial region \cite{32}, while the axisymmetric component can be 
model by the split-monopole (SM) magnetic field configuration \cite{34}. This composite magnetic field configuration has been used to investigated  magnetospheric topology, horizon-threading flux and particle dynamics around black holes \cite{33}.

For the SM magnetic filed configuration,  its non-zero vector potential component is given by \cite{34}
\begin{align}
A^{\rm SM}_{\phi} = -B_0\left|\cos\theta\right|,
\end{align}
where $B_0$ is a normalization constant. This configuration represents an axisymmetric poloidal magnetic field emerging from the northern hemisphere and entering the southern hemisphere.

For the external magnetic field, we employ the exact Bi\v{c}\'{a}k-Jani\v{s} solution \cite{35} to describe a uniform magnetic field inclined at an arbitrary angle with respect to the system symmetry axis. Therefore, the total four-vector potential $A_{\mu}$ is obtained by superposing the vector potentials of the two constituent magnetic fields,
\begin{align}
A_r &= -\frac{B_x}{2}(r-M)\sin 2\theta \sin\phi, \\
A_\theta &= -B_x\left(r^2\cos^2\theta - Mr\cos 2\theta\right)\sin\phi, \\
A_\phi &= -B_0\left|\cos\theta\right| + \frac{B_z}{2}r^2\sin^2\theta \notag\\
&\quad - \frac{B_x}{2}(r-M)r\sin 2\theta\cos\phi.
\end{align}
where $B_x$ and $B_z$ represent the horizontal and vertical components of the external magnetic field, respectively. The inclination angle $\alpha$ of the uniform external magnetic field with respect to the system symmetry axis is defined as
 $\alpha = \arctan B_x/B_z$. We set $B_0 = 1$ and keep the relative normalization of the SM and external components fixed at $B_0/\sqrt{B_x^2+B_z^2}=10$ for convenience.

Based on the total vector potential defined above, the magnetic field components in the static observer's frame (lab-frame) can be formulated as
\begin{subequations}
\begin{align}
B^r &= \frac{1}{\sqrt{-g}}\left(\frac{\partial A_\phi}{\partial\theta} - \frac{\partial A_\theta}{\partial\phi}\right), \\
B^\theta &= \frac{1}{\sqrt{-g}}\left(\frac{\partial A_r}{\partial\phi} - \frac{\partial A_\phi}{\partial r}\right), \\
B^\phi &= \frac{1}{\sqrt{-g}}\left(\frac{\partial A_\theta}{\partial r} - \frac{\partial A_r}{\partial\theta}\right), \\
B^t &= 0,
\end{align}
\end{subequations}
Here, $g$ is the Schwarzschild metric determinant. Following the ideal MHD relation \cite{36}, the laboratory-frame magnetic field components  $B^\mu$ can be transformed to the fluid-frame counterparts $b^\mu$ by
\begin{equation}
\begin{aligned}
b^t &= B^i u_i, \qquad
b^i &= \frac{B^i + b^t u^i}{u^t},
\end{aligned}
\end{equation}
where \(u^\mu\) denotes the fluid four-velocity.

\subsection{Polarimetric Observables}

Polarimetric observables are introduced below to quantitatively characterize the polarimetric properties of the synthetic images.  The net linear polarization fraction corresponds to the ratio of the net linear polarized intensity to the total Stokes $I$ intensity \cite{4,10},
\begin{equation}
m_{\rm net} = \frac{\sqrt{\left(\sum_i Q_i\right)^2 + \left(\sum_i U_i\right)^2}}{\sum_i I_i},
\end{equation}
which measures the global linear polarization degree after possible cancellation of polarized signals across  different image regions and therefore reflects the global coherence of the polarization structure. In addition,  the image-averaged linear polarization fraction \cite{4,10},
\begin{equation}
\langle |m| \rangle = \frac{\sum_i \sqrt{Q_i^2 + U_i^2}}{\sum_i I_i}.
\end{equation}
characterizes the average polarization level of the image. $\langle |m| \rangle$ is obtained by summing the polarized intensity of individual pixels across the resolved image and normalizing by the total intensity.

To further quantitatively characterize the polarization mode distribution, one can decompose the complex polarization field $\mathcal{P} = Q + iU$ into different azimuthal modes using Fourier expansion \cite{43,44},
\begin{equation}
\beta_m = \frac{1}{I_{\rm ann}} \int_{r_{\min}}^{r_{\max}} \int_0^{2\pi} \mathcal{P}(r,\phi)e^{im\phi} r\,d\phi\,dr,
\end{equation}
where \(r_{\min}\) and \(r_{\max}\) are chosen to encompass the entire polarized emission region. The chosen basis functions $ e^{im\phi}$ represent the possible spatial modes of EVPA variation with the azimuthal angle $\phi$ in the image plane, and $m$ is the corresponding mode order.  The quantity $I_{\rm ann}$ is the total intensity within the same integration region, which is given by
\begin{equation}
I_{\rm ann} = \int_{r_{\min}}^{r_{\max}} \int_0^{2\pi} I(r,\phi)r\,d\phi\,dr,
\end{equation}  
In general, the $\beta_m$ coefficients are complex numbers. The magnitude $|\beta_m|$ quantifies the degree of $m$-fold rotational symmetry in the EVPA distribution around the ring, while the argument $\angle\beta_m$ measures the average orientation of the symmetric component, relative to a reference pattern with vertical EVPA at image north. 
The mode $m=1$  mainly characterizes asymmetric patterns, while the mode $m = 2$ quantifies rotational symmetry signatures and serves as a robust discriminator between MAD and SANE flows for black hole images.

\begin{figure*}[htbp]
\centering
\includegraphics[width=1.0\textwidth]{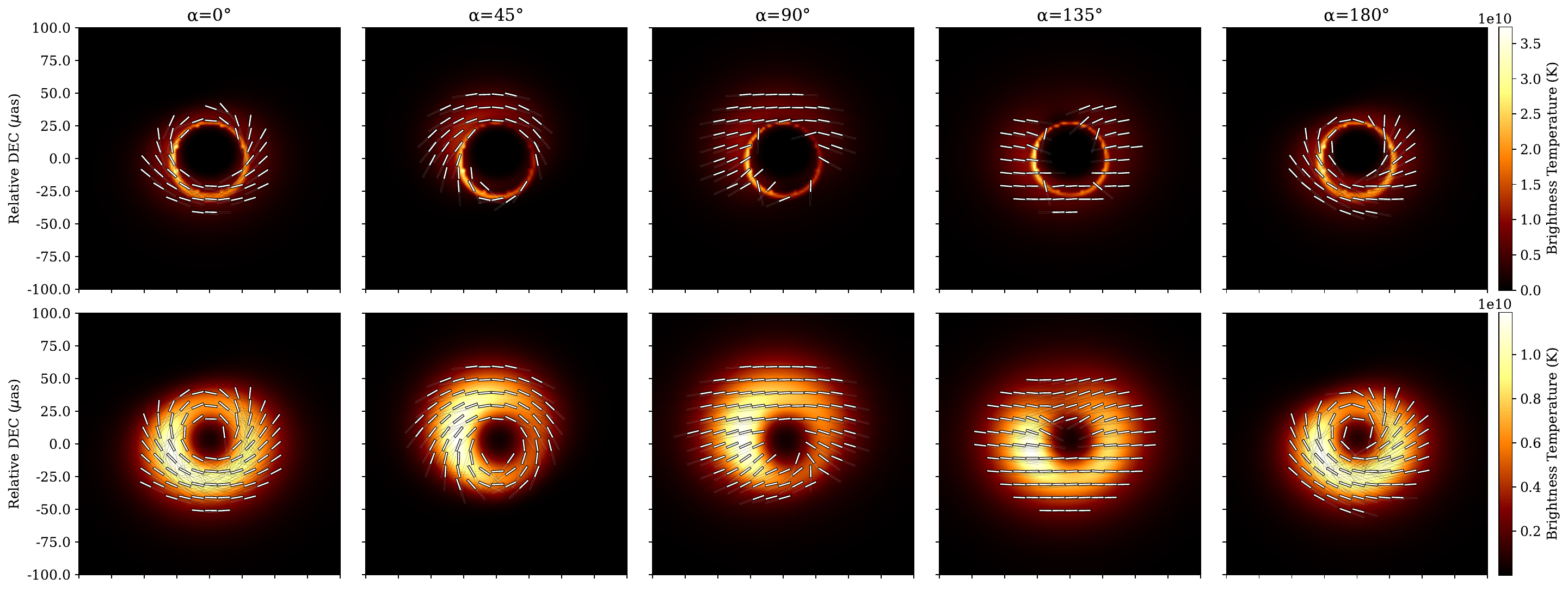}
\caption{Unblurred polarimetric images (top row)  and the corresponding blurred counterparts convolved with a $20 \mu\text{as}$ Gaussian kernel (bottom row) under tilted hybrid magnetic field configurations. For blurred images,  polarization ticks are shown in the region where $\mathcal{I} > 0.1\mathcal{I}_{\max}$ and $|\mathcal{P}| = \sqrt{\mathcal{Q}^2 + \mathcal{U}^2} > 0.2|\mathcal{P}|_{\max}$. Here, we set the observer azimuthal angle $\phi =0^{\circ}$ and its inclination $i=30^{\circ}$.}
\label{fig1}
\end{figure*}

\section{RESULTS}
\label{sec:results}

We now present the features of polarimetric images of a Schwarzschild black hole under a tilted hybrid magnetic field configuration. We begin by investigating how the inclination angle $\alpha$  of the external magnetic field modifies the macroscopic polarimetric morphology (\hyperref[sec:result-morphology]{Section III A}). Next, we explore the dependence of black hole images on the observer azimuthal angle for this hybrid magnetic configuration (\hyperref[sec:result-azimuth]{Section III B}). Finally, we quantitatively analyze the variations in polarimetric observables with the external magnetic field inclination angle  $\alpha$  (\hyperref[sec:result-observables]{Section III C}).

\subsection{Effects of External Magnetic Field Inclination on Polarimetric Morphology}
\label{sec:result-morphology}

Figure \ref{fig1} presents the polarimetric images for the fiducial model under five different external magnetic field inclinations for the fixed observer inclination $i=30^{\circ}$ and azimuthal angle $\phi =0^{\circ}$. As the magnetic field inclination $\alpha$ varies, significant changes appear in both the EVPA pattern and the brightness distribution. When $\alpha=0^\circ$ and $\alpha=180^\circ$, the external uniform magnetic field is respectively aligned and anti-aligned with the system symmetry axis, preserving a globally axisymmetric magnetic configuration after superposition. Consequently, the polarimetric images in these two limiting cases are overall similar, featuring a spiral-like EVPA pattern without the usual pure right- or left-handed direction. The deviation from the standard spiral distribution arises from that the reversal of the external magnetic field direction modify the total magnetic field distribution surrounding the black hole and thereby produce variations of the intrinsic emission in the accretion flow and the corresponding radiative transfer.
For the case $\alpha=45^\circ$ where the hybrid magnetic field deviates from axisymmetry, we find that the brightness distribution of the bright ring exhibits prominent asymmetry, but the EVPA presents a typical right-handed spiral azimuthal pattern, as reported in EHT observations.
With the magnetic inclination further increasing to $\alpha=90^\circ$, the spiral-like distribution of EVPA in the image plane undergoes gradual weakening, and a region dominated by horizontal EVPA arises in the upper portion of the image adjacent to the bright ring.
As the magnetic inclination  $\alpha=135^\circ$, the spiral-like EVPA structure is nearly absent and the EVPA is oriented almost entirely horizontally, except for a small patch at the top of the bright ring. Moreover, the brightness distribution along the bright ring is more dispersed than that for the  case with the external magnetic field inclination $\alpha=45^{\circ}$. The discrepancy in polarization images for external magnetic field inclinations of $45^{\circ}$ and $135^{\circ}$  can be attributed to the distinct total magnetic field configurations, given that polarization images are highly sensitive to magnetic field geometry.
After being blurred with a $20\ \mu\text{as}$ circular Gaussian, we find that some fine structures in the polarization images are smoothed, yet the dependence of the overall EVPA patterns on the magnetic field inclination remains clearly discernible. The brightness distribution also exhibits a clear dependence on the external magnetic field inclination \(\alpha\), with the primary bright region shifting around the emission ring. These results indicate that the external magnetic field inclination significantly influences the observed polarimetric morphology through modifying the hybrid magnetic field configurations.

\begin{figure*}[!t]
\centering
\includegraphics[width=1.0\textwidth]{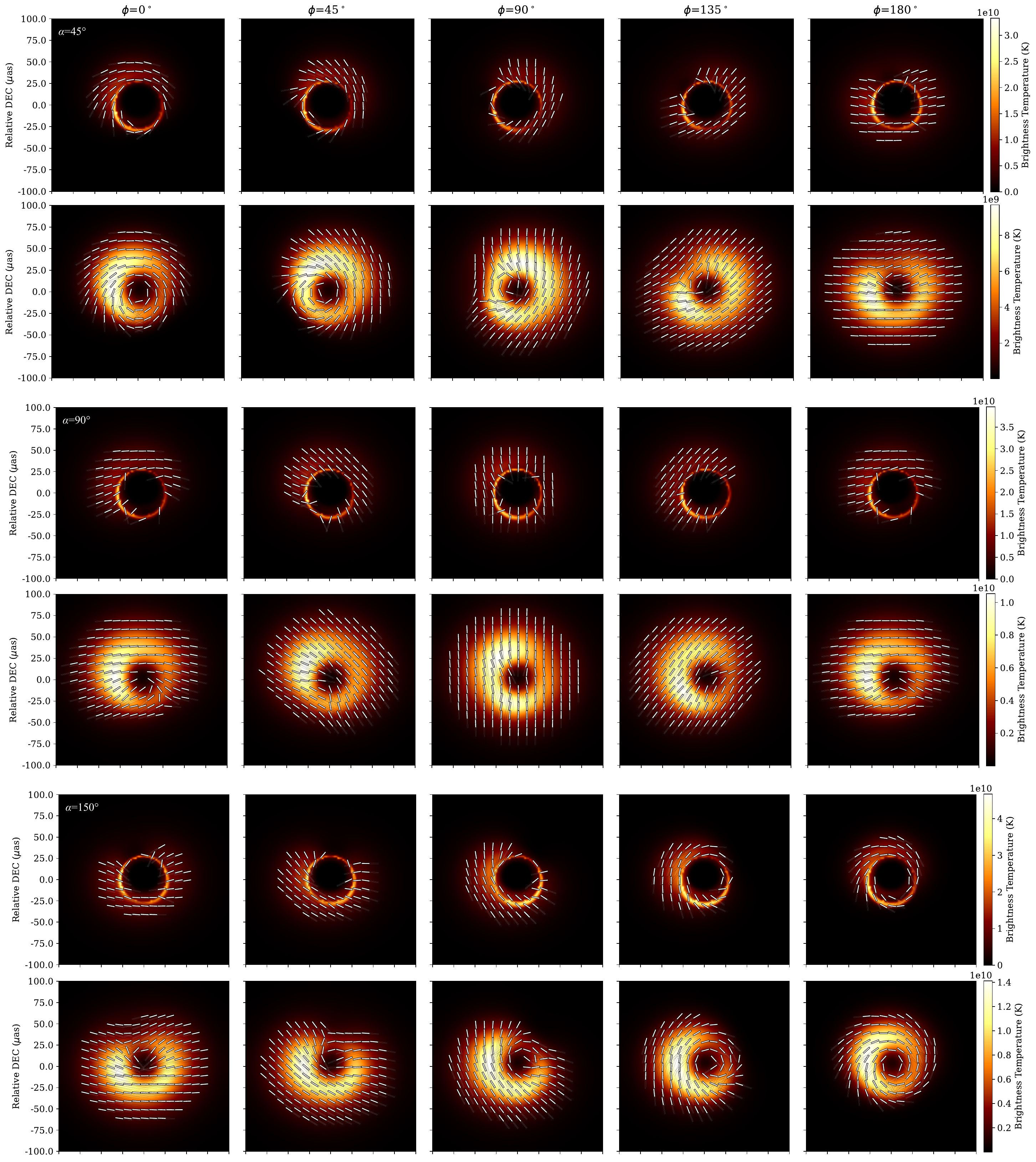}
\caption{Polarimetric images of the fiducial model for three fixed external magnetic field inclination angles of $\alpha=45^\circ$, $90^\circ$ and $150^\circ$ under different observer azimuthal angles. All other model configurations and plotting criteria are the same as those described in Fig 1.}
\label{fig2}
\end{figure*}

\subsection{Azimuthal Dependence of Polarization images}
\label{sec:result-azimuth}

Here we consider three cases of external magnetic field inclinations $\alpha=45^\circ$, $90^\circ$ and $150^\circ$, with a fixed observer inclination $i=30^\circ$, and probe the impact of the observer's azimuthal angle $\phi$ on polarimetric morphology. In Figure \ref{fig2}, we find that for the external magnetic field inclination $\alpha=45^\circ$, 
the EVPA pattern still exhibits a right-hand-spiral-like structure when the observer's azimuthal angle $\phi$ increases from $0^{\circ}$ to $45^{\circ}$. With increasing observer's azimuthal angle up to $90^{\circ}$, the spiral-like structure in polarization images vanishes, and the EVPA distribution becomes horizontally symmetric. With further increase of the observer’s azimuthal angle  $\phi$, EVPA directions across the image gradually become uniform, and the EVPA distribution is oriented nearly horizontally at $\phi=180^\circ$ apart from a small central region. 
For an external magnetic field inclination $\alpha=90^\circ$, we find that EVPA directions across the image are nearly identical and the EVPA orientation rotates counterclockwise as the observer's azimuthal angle $\phi$ gradually increases. For an external magnetic field inclination $\alpha=150^\circ$, we find that the EVPA distribution evolves from nearly uniform orientations into a right-handed spiral-like pattern, which is opposite to that in the case $\alpha=45^\circ$. These results show a novel dependence of polarization images on the observer's azimuthal angle $\phi$
due to the breaking of axisymmetry of the total hybrid magnetic field originating from  the tilted external magnetic field component. Moreover, for the hybrid magnetic field configuration, we also find in blurred images a more complex brightness distribution featuring multiple localized bright spots rather than a single dominant bright spot at certain observer's azimuthal angles. The distinctive signatures of tilted magnetic fields encoded in these black hole images can be exploited to probe non-axisymmetric magnetic components around the black hole. 
\begin{figure*}[!t]
\centering
\includegraphics[width=\textwidth]{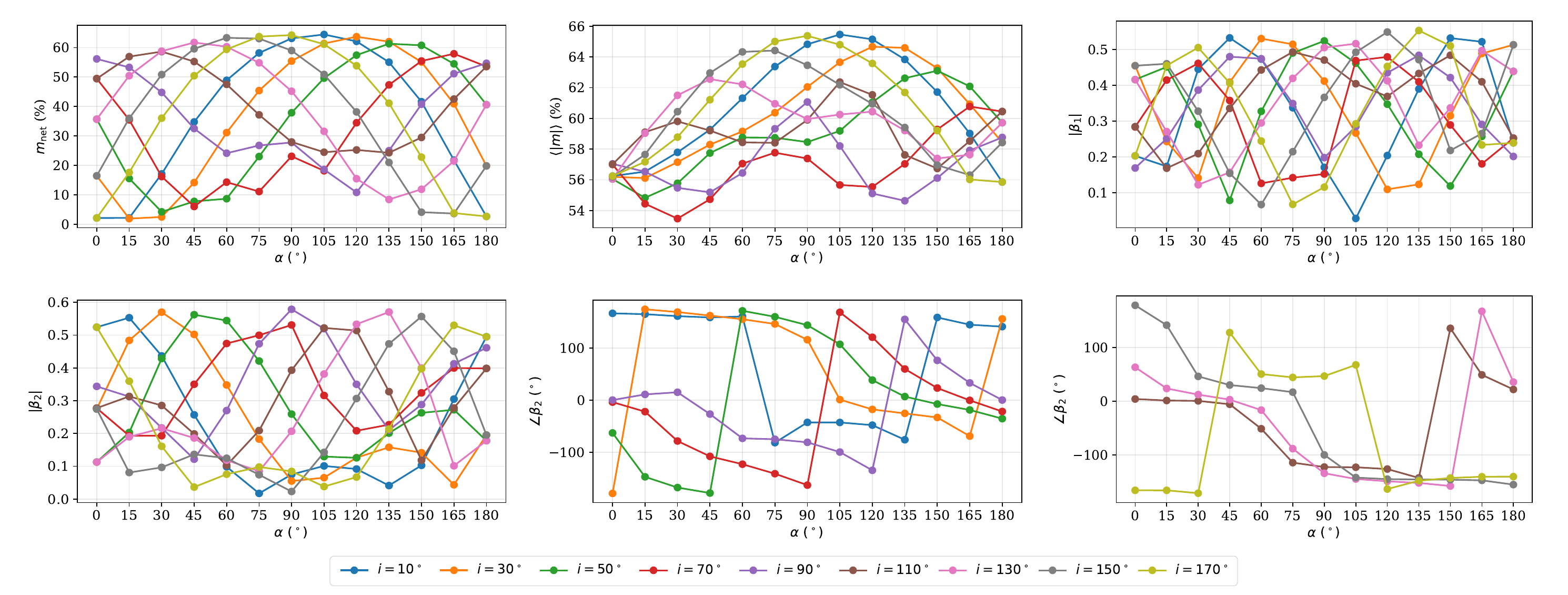}
\caption{Dependence of polarimetric observables on the external magnetic field inclination angle $\alpha$ for different observer inclinations at the given observer azimuth $\phi=0^\circ$.}
\label{fig:polarization-observables}
\end{figure*}

\subsection{Polarimetric Signatures}
\label{sec:result-observables}

Figure \ref{fig:polarization-observables} shows change of polarimetric observables with the external magnetic field inclination \(\alpha\) for different observer inclinations, at a fixed observer azimuth angle \(\phi=0^\circ\). 
With increasing external magnetic field inclination \(\alpha\), the net linear polarization fraction \(m_{\rm net}\) behaves differently across observer inclinations: it first increases and then decreases for \(i=10^\circ\); for \(i=30^\circ\)-\(90^\circ\), it undergoes an initial decrease, a subsequent rise, and a final drop; for \(i=110^\circ\)-\(150^\circ\), it increases first, decreases, and increases once more; and for \(i=170^\circ\), it increases initially before decreasing. Moreover, the \(\alpha\) value at which the net linear polarization fraction reaches its maximum shifts rightward with increasing observer inclination. Compared with $m_{\rm net}$, $\langle |m| \rangle$ remains within a relatively narrow range for all observer inclinations and exhibits a broadly similar but smoother overall variation, because it is less sensitive to cancellation between polarization components from different image regions. In addition, the peak values of both linear polarization fractions are distributed approximately symmetrically about \(i=90^\circ\). 

The azimuthal polarization modes exhibit a more complex dependence on the external magnetic field inclination. The quantities  $|\beta_1|$ and $|\beta_2|$ varies significantly with $\alpha$, but no systematic trend is found across different observer inclinations.  
We find that the quantity \(|\beta_2|\) reaches its maximum when the external magnetic field inclination \(\alpha\) approximately equals the observer inclination $i$, i.e., \(\alpha \simeq i\), which means that the observed polarimetric image exhibits the apparent rotationally symmetric feature as the observer is aligned with the uniform external magnetic field.
The variation of phase \(\angle\beta_2\) with the inclination \(\alpha\) of the external magnetic field is highly complex, with no discernible dependence. The apparent discontinuities in the curves \(\angle\beta_2-\alpha\) do not imply abrupt physical changes in the polarization pattern because
these jumps mainly occur when \(\beta\) transitions from near \(-180^\circ\) to near \(180^\circ\).

\begin{figure*}[!t]
\centering
\includegraphics[width=\textwidth]{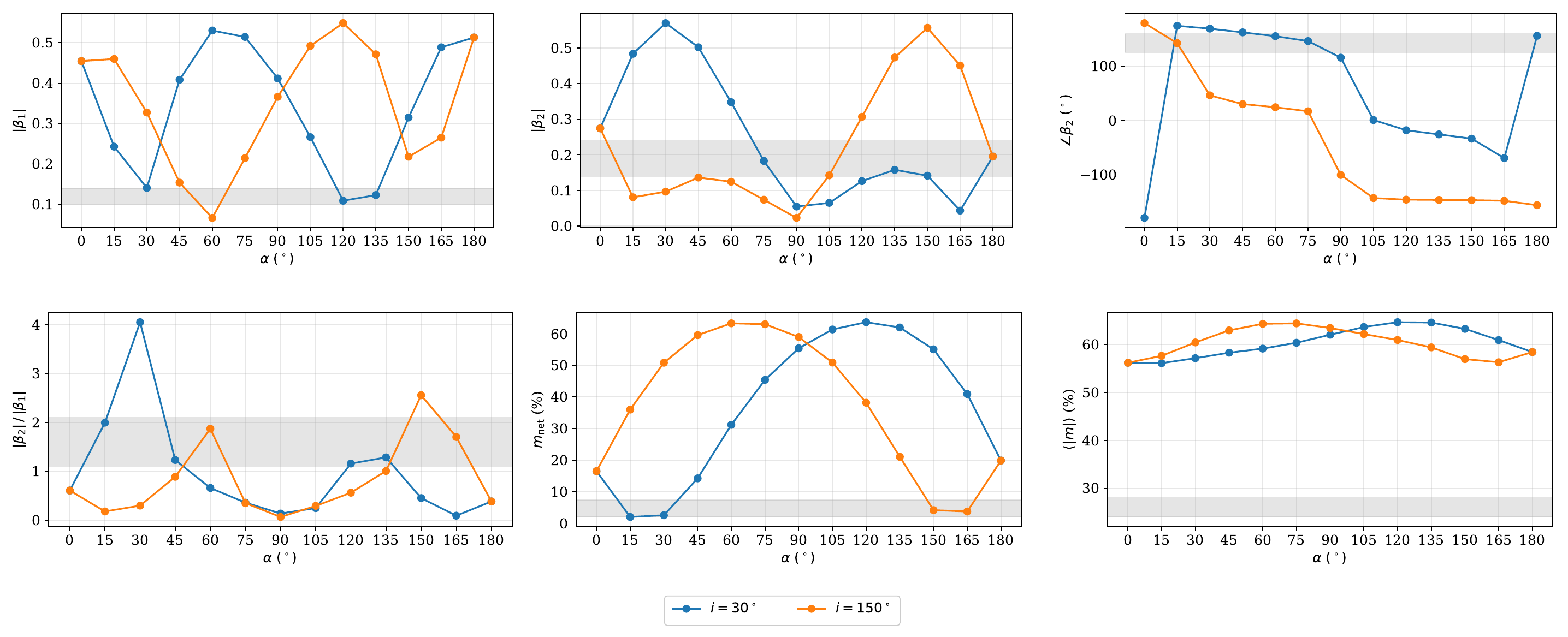}
\caption{Image-domain polarimetric observables as functions of the external magnetic field inclination angle $\alpha$ for observer inclinations $i=30^\circ$ and $150^\circ$. The gray shaded regions indicate the corresponding EHT observational constraint ranges.}
\label{fig:observational-constraints}
\end{figure*}

\section{OBSERVATIONAL CONSTRAINTS}
\label{sec:observational-constraints}
\fontsize{11}{13}\selectfont

To check whether the tilted hybrid magnetic field model is compatible with the polarimetric properties of Sgr A*, we compare the polarimetric observables with the data from the EHT \cite{10}. Based on the parameters of the fiducial model presented in \hyperref[sec:results]{Section III}, we here focus on only two representative observer inclinations, \(i=30^\circ\) and \(150^\circ\). 
\hyperref[fig:observational-constraints]{Figure 4} shows the constraint imposed by \(|\beta_1|\) yields permitted ranges of the external magnetic field inclination \(\alpha\) of \([115^\circ,137^\circ]\) for \(i=30^\circ\) and \([47^\circ,70^\circ]\) for \(i=150^\circ\). By contrast, the \(\beta_2\) mode constraint gives allowed \(\alpha\) ranges of \([70^\circ,80^\circ]\cup[170^\circ,180^\circ]\) at \(i=30^\circ\) and \([5^\circ,15^\circ]\) at \(i=150^\circ\). Furthermore, the constraint from the net linear polarization fraction \(m_{\rm net}\) restricts \(\alpha\) to \([13^\circ,40^\circ]\) for \(i=30^\circ\) and \([142^\circ,168^\circ]\) for \(i=150^\circ\).
This implies that there exists no overlapping region between the permitted parameter spaces for \(\alpha\) obtained from the three separate constraints imposed by modes $m=1$, $m=2$ and the net linear polarization fraction \(m_{\rm net}\) for both observer inclinations. Although the parameter regions permitted by $|\beta_2|/|\beta_1|$ overlap with those from \(m_{\rm net}\), the constraint from the image-averaged linear polarization fraction $\langle |m| \rangle$  disfavors this tilted hybrid magnetic field model composed of an axisymmetric SM magnetic field and a tilted uniform external magnetic field. This incompatibility among constraints derived from multiple polarimetric observables may originate from the following two aspects. First, the magnetic fields in the vicinity of real black holes may not possess such simple geometries and are likely much more intricate than assumed in this model. 
 Second, we model the accretion disk using a semi-analytic RIAF model, which is not sufficiently accurate to fully characterize the properties of the black hole accretion disk. Constraining the tilted hybrid magnetic field configuration through polarimetric observables may require higher-precision GRMHD simulations of the accretion flow surrounding the black hole.

\section{Conclusions}
\label{sec:conclusions}
\fontsize{11}{13}\selectfont
In this work, we investigate the polarimetric signatures of Sgr A* using a stationary semi-analytical RIAF model with a tilted hybrid magnetic field configuration. The hybrid magnetic field is constructed by superposing an axisymmetric split-monopole field with an inclined uniform external magnetic field. We carry out general-relativistic polarized radiative transfer simulations to explore how the inclination of the external magnetic field affects the polarimetric morphologies of black hole images and the associated polarimetric observables. By comparing with EHT observational data, we place constraints on such tilted hybrid magnetic field configurations around Sgr A*. 

\begin{itemize}
\item The tilted external magnetic field may break the global axisymmetry of the hybrid magnetic field configuration, yielding polarized black hole images that deviate from the conventional purely right- or left-handed polarization patterns. With increasing external magnetic field inclination from $0^\circ$ to $135^\circ$, the spiral-like polarization patterns gradually vanish, and the EVPA becomes nearly purely horizontal. As the inclination further increases to $180^\circ$, these spiral-like patterns gradually reappear. The brightness distribution also exhibits a clear dependence on the external magnetic field inclination, with the primary bright region shifting around the emission ring. 
\end{itemize}

\begin{itemize}
\item The tilted external magnetic field leads to a novel azimuthal dependence of polarization images on the observer angle \(\phi\). As the azimuthal angle varies, the corresponding polarization images for a fixed observer inclination transition between spiral-like patterns and nearly uniform distributions. Moreover, in blurred images, the brightness distribution becomes more complex at certain observer azimuthal angles, featuring multiple localized bright spots instead of a single dominant bright spot. The distinctive signatures of tilted magnetic fields encoded in these black hole images can be exploited to probe non-axisymmetric magnetic components around the black hole.
\end{itemize}

\begin{itemize}
\item The azimuthal polarization modes exhibit a more complex dependence on the external magnetic field inclination. 
The quantity \(|\beta_2|\) reaches its maximum for observer azimuth \(\phi=0^\circ\) when the external magnetic field inclination \(\alpha\) is approximately equal to the observer inclination $i$, i.e., \(\alpha \simeq i\). This implies that the observed polarimetric image displays an apparent rotationally symmetric feature when the observer is aligned with the uniform external magnetic field. Variations in the net linear polarization fraction \(m_{\rm net}\) with the external magnetic field inclination depend strongly on the observer inclination. With increasing observer inclination, the peak position of \(m_{\rm net}\) shifts in the direction of increasing \(\alpha\). In addition, the peak values of both linear polarization fractions \(m_{\rm net}\) and $\langle |m| \rangle$ are distributed approximately symmetrically about \(i=90^\circ\). 
\end{itemize}

\begin{itemize}
\item A comparison with EHT polarimetric constraints on Sgr A* indicates that the allowed parameter regions obtained from multiple polarimetric observables are non-overlapping. The resolution of such incompatibility may rely on improvements to the magnetic field model around the black hole and the adoption of higher-precision GRMHD simulations of accretion disks, rather than semi-analytic RIAF models.

\end{itemize}

The external magnetic field inclination substantially modifies the polarization patterns and brightness distributions in black hole images. The non-axisymmetry induced by the tilted magnetic field component makes the polarization morphology dependent on the observer azimuthal angle. Constraints from EHT data suggest that, besides the large-scale tilted external magnetic field geometry, more intricate magnetic field configurations and more complete accretion-disk models are required.

%%%%%%%%%%%%%%%%%%%%%%%%%%%%%%%%%%%%%%%%%%%%%%%%%%%%%%%%%%%%%%%%%%%%%%%%%%%%
\begin{acknowledgments}
This work was supported by the National Natural Science Foundation of China under Grant No.12675066, 12275078, 11875026, 12035005, 2020YFC2201400 and the innovative research group of Hunan Province under Grant No. 2024JJ1006.
\end{acknowledgments}

\noindent{Conflict of Interest}\quad The authors declare that they have no conflict of interest.
%%%%%%%%%%%%%%%%%%%%%%%%%%%%%%%%%%%%%%%%%%%%%%%%%%%%%%%%%%%%%%%%%%%%%%%%%%%%
\bibliography{imagetiltedblackhole}
\bibliographystyle{sn-aps}
%%%%%%%%%%%%%%%%%%%%%%%%%%%%%%%%%%%%%%%%%%%%%%%%%%%%%%%%%%%%%%%%%%%%%%%%%%%%
\end{document}